# Application of Quaternions to Obtain Analytic Solutions to Systems of Polarization Components

Michael G. Taylor
Atlantic Sciences LLC, Laurel, Maryland, USA
mtaylor@atlanticsciences.com

## Article info

Keywords:
Fiber optics
Optical fiber communication
Optical retarders
Polarization
Quaternions
Waveplates

## Abstract

This paper describes the passage of light through a system of waveplates mathematically in terms of quaternions, an extension of the complex numbers, instead of the more usual Jones vectors and Jones matrices. Both the light beam and the waveplate are represented by a quaternion. It is possible to manipulate the quaternion expression more readily than the Jones matrix-vector expression; for example it can be inverted. The quaternion form of a waveplate is compactly related to its retardance and fast/slow axes, and the quaternion of a signal is closely related to its state of polarization (SOP), either expressed as a vector on the Poincaré sphere or as a polarization ellipse. The paper presents rules to decide if two optical signals are aligned or orthogonal in phase or in polarization from their quaternions, and presents the quaternion operations to change the phase or change the SOP. Several mathematical tools are identified, such as partial conjugation, to rearrange a quaternion expression, a tricky operation because multiplication does not commute. Put together, these advances let us understand how waveplates can act on a light beam to produce desired behavior. Finally, the quaternion math is put to work on two problems. A new endless optical phase shift system is designed out of waveplates. A prior solution to the problem used five waveplates, and in this paper the same task is done with only three waveplates. Also, failures of a polarization controller are studied, and found to be caused by singularities, which can occur frequently.

# 1 Introduction

## 1.1 Systems of polarization components

The conventional way to deal mathematically with coherent light passing through passive optical components is by using Jones algebra [1, 2]. The signal is described by a Jones vector (a 2x1 complex vector) and the optical component by a Jones matrix (a 2x2 complex matrix). For problems in the forward direction, such as calculating the output state of polarization resulting from a given input and given optical components, the Jones algebra directly gives the desired result. However, when calculating the optical components that give a specified input and output, the Jones algebra can certainly state the problem, but may not be helpful in arriving at an analytic solution. This kind of reverse direction problem is often solved numerically.

The Jones matrix, which has eight degrees of freedom, is able to include behavior of a waveplate and polarization dependent loss and optical phase shift. When the optical component is a waveplate it is already known that a quaternion, having four degrees of freedom, is a more compact way to describe it than a Jones matrix [3]. This paper goes a step further by also representing the optical signal as a quaternion. The whole problem is then written in terms of quaternions, without Jones vectors or matrices. It is an improvement on the Jones algebra in that reverse direction-type problems can be solved analytically as well as the forward direction. For example Section 4 shows how quaternion division solves a simple reverse direction problem, and Section 7 derives analytically the algorithm to control a new phase shifter design.

The state of polarization (SOP) of a signal is built into the Jones vector. There are two other ways that are often used to represent the SOP, as a vector on the Poincaré sphere, or as a polarization ellipse. The latter is more a diagrammatic than a functional representation. When we move to the quaternion to represent the optical signal, a simple mathematical connection is found to all these SOP representations: the Jones vector (equation (8)), the Poincaré sphere (equation (10)), and the polarization ellipse (equation (11)).

In a similar way, there is a simple relationship (equation (17)) between the quaternion representing a waveplate and its basic parameters, the fast/slow axes and the retardance. This means that, unlike the Jones algebra version, the quaternion equation describing light passing through a system of waveplates directly connects the parameters of the signal, the SOP and phase, and the parameters of the waveplates. The equation is ready to be solved for the desired (unknown) parameters in terms of the known parameters. Such a quaternion expression can appear difficult to manipulate, because quaternion multiplication is not commutative. Some mathematical tools have been identified that help solve a quaternion equation: the partial conjugate in Section 3.4 and four Multiplication Reordering Rules in Section 3.5. These tools are put to use in several places in the paper to show how a desired parameter can be extracted from a quaternion expression.

The final worked examples employ the new quaternion treatment. In Section 7 a previously unsolved problem, an endless phase shifter comprising only three waveplates, is written in terms of quaternions, and then solved analytically. The orientations of the waveplates are derived that produce the desired phase shift. Section 8 uses quaternion algebra to purposely look for cases where a polarization controller fails, and finds that this condition, a singularity, often occurs.

## 1.2 Quaternion basics

The quaternions are an extension of the complex numbers that were discovered by William Rowan Hamilton in 1843 [4, 5]. Instead of one imaginary number there are three of them, $i$, $j$ and $k$, that are different from one another. They follow the non-commutative multiplication rules

$$\begin{gathered} i^2 = j^2 = k^2 = ijk = -1 \\ ij = -ji = k \qquad jk = -kj = i \qquad ki = -ik = \mathrm{j} \end{gathered} \tag{1}$$

A quaternion $q$ is defined by four real coefficients $q_0..q_3$

$$q = q_0 + q_1 i + q_2 j + q_3 k$$

The first component is known as the scalar part, denoted by Sc(·), and the remainder as the vector part Ve(·).

$$\mathrm{Sc}(q) = q_0 \qquad \mathrm{Ve}(q) = q_1 i + q_2 j + q_3 k$$

A quaternion with no scalar part is known as a vector quaternion (or a pure quaternion).

Addition of quaternions is done by simply adding the respective coefficients. Multiplication of two quaternions $p$ and $q$ follows the product rules in (1), and can be summarized as

$$pq = p_0 q_0 - \mathrm{Ve}(p) \cdot \mathrm{Ve}(q) + p_0\,\mathrm{Ve}(q) + q_0\,\mathrm{Ve}(p) + \mathrm{Ve}(p) \times \mathrm{Ve}(q)$$

where in this equation the dot symbol refers to the conventional dot (scalar) product of two vectors, and the × symbol refers to the cross (vector) product.

The quaternion conjugate, denoted by superscript †, has the vector part negated.

$$q^\dagger = q_0 - q_1 i - q_2 j - q_3 k$$

The absolute value (or norm) of the quaternion is defined by

$$|q| = \sqrt{q^\dagger q} = \sqrt{q q^\dagger} = \sqrt{{q_0}^2 + {q_1}^2 + {q_2}^2 + {q_3}^2}$$

A quaternion with absolute value 1 is known as a unit quaternion. For any unit vector quaternion $v$

$$v^2 = -1 \tag{2}$$

Quaternion division is the inverse of multiplication, and therefore is also not commutative. Two division signs are used, / for right division and \ for left division. For quaternions $p$ and $q$

$$p/q = \frac{pq^\dagger}{|q|^2} \qquad q\backslash p = \frac{q^\dagger p}{|q|^2}$$

The four quaternion base components can be associated with the following 2x2 complex matrices

$$\begin{gathered} 1 \to \begin{pmatrix} 1 & 0 \\ 0 & 1 \end{pmatrix} \qquad i \to \begin{pmatrix} h & 0 \\ 0 & -h \end{pmatrix} \\ j \to \begin{pmatrix} 0 & -1 \\ 1 & 0 \end{pmatrix} \qquad k \to \begin{pmatrix} 0 & h \\ h & 0 \end{pmatrix} \end{gathered} \tag{3}$$

where $h$ is the conventional imaginary number, the square root of real number -1. It can be verified that the choice of basis set (3) produces the products seen in (1) given that the order of matrix multiplication is reversed. (In this quaternion treatment a sequence of optical elements is purposely written from left to right, in the order light passes through them. This is the opposite direction to Jones matrix multiplication, which goes from right to left.)

## 1.3 Prior work using quaternions in optical transmission

The fiber optic communications industry is four decades old and the modern mathematical treatment of polarization began in the 1940s. Unsurprisingly some of the mathematical ideas in this paper, the application of quaternions, have already appeared in the literature. The complete set of relationships between quaternions and the optical variables presented here and the tools to use the quaternions have not been reported before. This section will try to go over what features of this paper are new and what have already been discovered.

It has been noticed by several authors, sometimes independently, that the multiplication of quaternions is isomorphic to the multiplication of Jones matrices representing waveplates, which means we can express waveplate concatenation as quaternion multiplication [3, 6, 7, 8, 9, 10]. Some authors have expressed the relationship in terms of Pauli spin matrices instead of quaternions [11, 12, 13, 14, 15, 16]. The Pauli spin matrices are closely related to the quaternions. In the basis set of (3)

$$\sigma_1 = -hk \qquad \sigma_2 = hj \qquad \sigma_3 = -hi$$

While a quaternion has been used to represent an optical signal before, for example Pellat-Finet used it like a Stokes vector to describe partially polarized light [11], this paper appears to be the first time a quaternion is used to represent the electric field directly (equation (8)), the equivalent of a Jones vector.

Most of the quaternion manipulations of Section 3 are well known, but the author has not found mention of partial conjugation, or the four reordering rules collected in one place. The result (11) that the quaternion is equal to the product of the exponentials of the polarization ellipse parameters is new. The relations in Section 4 are new that modify optical parameters of a signal via the quaternions (changing the phase, going to orthogonal SOP) and identify optical orthogonality conditions from two quaternions. An equivalent formula to equation (10), giving the Stokes vector of a signal from its quaternion, has been found by workers using the Pauli spin matrix basis [17, 11]. That formula, the inner product of the Jones vector with the spin vector, was simply presented as a valid equation, while the quaternion treatment here explains where (10) comes from.

The central result of Section 5, equation (17), identifies the Euler parameter of the quaternion representing a waveplate as the retardance–Stokes vector product. The Pauli spin matrix papers include an exponential (or a sum of cosine and sine terms) that comes from the underlying relationship behind (17), but they do not identify the exponential as Euler's formula and it contains an imaginary number that arises because the Pauli spin matrices are not closed under multiplication.

The expression in Section 6 to enact a polarization dependent loss element with quaternions is new. Finally, the worked example of Section 7 is new, how to produce an endless phase shift from only three waveplates.

## 1.4 Quaternion & complex number notation

This paper uses the notation Re(·), Im(·), arg(·) and the asterisk superscript $(\cdot)^*$ to refer to the real part, imaginary part, argument (angle) and complex conjugate respectively of a complex quantity. As already mentioned, for quaternions the scalar and vector parts and the quaternion conjugate are denoted by Sc(·), Ve(·) and $(\cdot)^\dagger$. Sometimes we have a quaternion quantity which has only one vector component, *i* or *j* or *k*, and it can be a useful tool to treat is as complex number. The complex notation Re(·) etc. will be applied only to such complex-like quaternions.

## 2 Waveplates

A waveplate (also known as a retarder or retardation plate or birefringent element) is an optical component whose propagation delay depends on the input state of polarization. Light launched on one SOP (the slow axis) experiences a greater phase shift than light launched on the orthogonal SOP (the fast axis). The difference between the phase shifts, at a specific wavelength, is the retardance $2\eta$ (also known as the retardation). This paper considers only waveplates where the fast axis and slow axis phase shifts have equal magnitude and opposite signs. A more general waveplate is equivalent to a balanced waveplate followed or preceded by a polarization-independent phase shift. Typically the extra phase shift is of no consequence. For the phase shifter of Section 7, for example, the solution still sweeps the desired $2\pi$ of phase even if waveplates with asymmetric fast and slow phase shifts are used.

The transformation of a waveplate can be represented by a Jones matrix which has a symmetry to its coefficients. Given $a$ and $b$ are complex numbers the Jones matrix has the form

$$\begin{pmatrix} a & -b^* \\ b & a^* \end{pmatrix} \tag{4}$$

When the matrix has a determinant of 1 it belongs to a subset of the unitary matrices. The class of matrix (4) is closed under multiplication and addition, and multiplication is not commutative, so it forms a division ring. The four matrices associated with the quaternion elements in (3) also belong to this class of matrix. That means any Jones matrix representing a waveplate can be replaced by a quaternion $p$, whose four coefficients $p_0..p_3$ are the weights of the four matrices in (3). Some examples of waveplates and their quaternion representation are given in Table 1.

**Table 1** Waveplate transformations and their quaternions

| quaternion | transformation |
|---|---|
| $i$ | half waveplate, slow axis horizontal |
| $\frac{1}{2}\sqrt{2}(1+i)$ | quarter waveplate, slow axis horizontal |
| $\frac{1}{2}\sqrt{2}(1-i)$ | quarter waveplate, fast axis horizontal |
| $j$ | physical rotation through $\pi/2$, or half waveplate with circular fast/slow states |

According to Jones calculus, when two waveplates are concatenated their combined effect is expressed by multiplying their Jones matrices. In quaternion terms the equivalent is obtained by multiplying the two quaternions associated with the two waveplates.

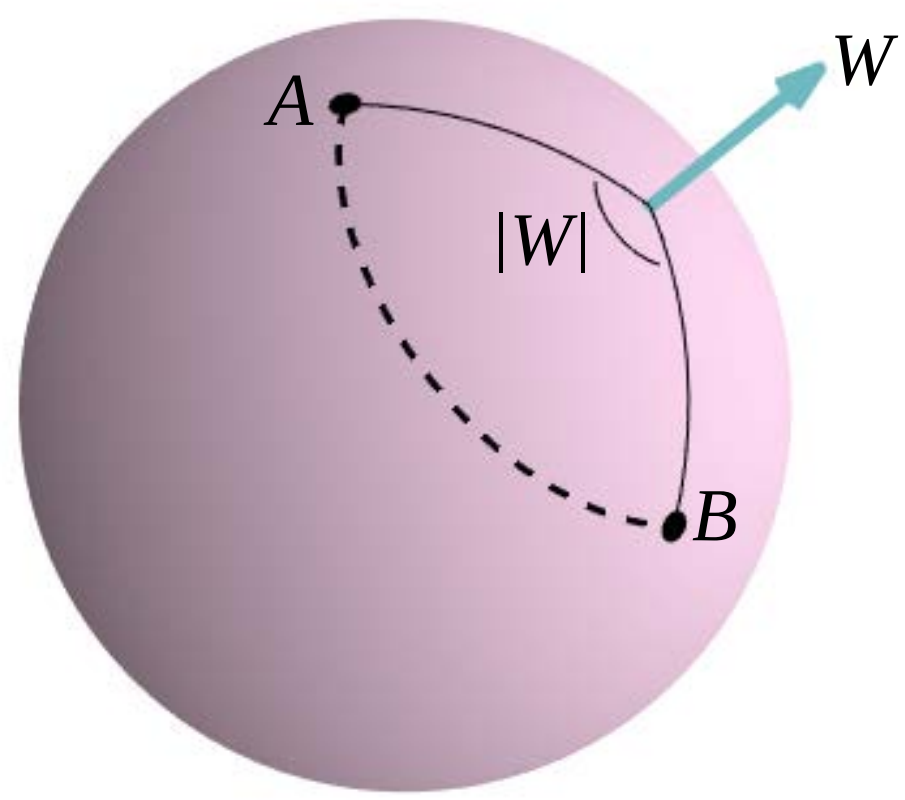

**Fig. 1.** The effect of a waveplate seen on the Poincaré sphere. The waveplate transforms the SOP from $A$ to $B$, a rotation about the axis of vector $W$ by angle $|W|$.

A waveplate can also be characterized by its effect on the SOP of a signal seen on the Poincaré sphere. The output Stokes vector is equal to the input Stokes vector rotated (precessed) about an axis corresponding to the fast-slow states of the waveplate (Fig. 1). The angle of precession is equal to the waveplate's retardance $2\eta$. The transformation can be summarized as a vector $W$ in Stokes space, whose direction is that of the slow axis and whose length is the retardance. The utility of vector $W$ is limited, however, in that the result of two waveplates in sequence is not equal to the sum of the two Stokes space vectors. Instead, one of the vectors has to be precessed around the other before the two are added, as described in [18].

# 3 Further quaternion math

Section 1.2 of the introduction included the mathematical properties of quaternions to understand how they are used to represent waveplates. The new application of quaternions in this paper requires some more mathematical manipulations.

## 3.1 Orthogonality

Two quaternions $p$ and $q$ are said to be orthogonal if

$$\mathrm{Sc}(p^\dagger q) = \mathrm{Sc}(pq^\dagger) = p_0q_0 + p_1q_1 + p_2q_2 + p_3q_3 = 0$$

Any quaternion $qv$ or $vq$ is orthogonal to $q$, and $pvq$ is orthogonal to $pq$, where $v$ is a vector quaternion.

## 3.2 Euler's theorem

Euler's theorem for complex numbers can be extended to quaternions. For a unit vector quaternion $v$ (which is a square root of -1 according to (2)) and a scalar $\theta$

$$e^{v\theta} = \cos\theta + v\sin\theta$$

This relation can easily be extended to a definition for exponential and logarithm of a general quaternion. Note that the quaternion exponential is subject to non-commutative multiplication. For quaternions $p$ and $q$

$$e^p e^q \not\equiv e^q e^p \not\equiv e^{p+q}$$

In the special case where $\mathrm{Ve}(p)$ is parallel to $\mathrm{Ve}(q)$

$$e^p e^q = e^q e^p = e^{p+q}$$

## 3.3 Precession

An expression of the form

$$p = e^{-v\theta}\, q\, e^{v\theta}$$

where $q$ is a quaternion, $v$ is a unit vector quaternion and $\theta$ is a scalar, means $p$ is a precessed version of $q$. The scalar parts of $p$ and $q$ are the same. The vector part of $p$ is the vector part of $q$ precessed by angle $2\theta$ around axis $v$.

## 3.4 Partial conjugation

The partial conjugate of a quaternion involves inverting one or two components of the vector part only. The single partial conjugate with respect to unit vector quaternion $v$ is denoted by superscript $\dagger v$. For example

$$q^{\dagger i} = q_0 - q_1 i + q_2 j + q_3 k$$

$q+q^{\dagger v}$ has no $v$-component, and $q- q^{\dagger v}$ contains only a $v$-component. Importantly, both $qq^{\dagger v}$ and $q^{\dagger v}q$ (which are different from one another) contain no $v$-component. The $v$-conjugate can be expressed as (see Appendix B)

$$q^{\dagger v} = -vq^{\dagger}v \tag{5}$$

For two quaternions $p$ and $q$ the product follows the rule

$$(pq)^{\dagger v} = q^{\dagger v}p^{\dagger v} \tag{6}$$

The double quaternion conjugate, where two mutually orthogonal vector components are inverted, is denoted either by superscript $\dagger$ followed by the two negated components or by superscript $\dagger\dagger$ followed by the non-negated component. For example

$$q^{\dagger ij} = q^{\dagger\dagger k} = q_0 - q_1 i - q_2 j + q_3 k$$

$q+q^{\dagger\dagger v}$ has only scalar and $v$-components, while $q-q^{\dagger\dagger v}$ has no scalar or $v$-components. Unlike single partial conjugation, the products $qq^{\dagger\dagger v}$ and $q^{\dagger\dagger v}q$ do not have a particular vector component absent. The double conjugate can be expressed as

$$q^{\dagger\dagger v} = -vqv \tag{7}$$

The product rule for double conjugation does not involve exchanging the terms like single conjugation.

$$(pq)^{\dagger\dagger v} = p^{\dagger\dagger v}q^{\dagger\dagger v}$$

The triple conjugate $q^{\dagger ijk}$ is of course the same as the full quaternion conjugate $q^{\dagger}$ already presented in Section 1.2. The product rule for the full conjugate is

$$(pq)^{\dagger} = q^{\dagger}p^{\dagger}$$

These different forms of quaternion partial conjugate will let us express relationships between optical quantities, and also act as a tool to aid in solving quaternion equations.

### 3.5 Multiplication reordering rules

One of the challenges of quaternion algebra is that because multiplication is not commutative it is hard to cancel two opposing terms that are not adjacent in a product. The four rules below have identified special cases where the order of terms can be rearranged.

*Multiplication Reordering Rule 1:* If $p$ and $q$ have vector parts that are parallel, i.e. $\mathrm{Ve}(p) = \lambda\,\mathrm{Ve}(q)$ where $\lambda$ is a scalar, then

$$pq = qp$$

*Multiplication Reordering Rule 2:* If vector quaternion $v$ is orthogonal to quaternion $q$ then

$$qv = vq^{\dagger}$$

*Multiplication Reordering Rule 3:* If $u$ and $v$ are mutually orthogonal unit vector quaternions (which means $uv$ is also a unit vector quaternion orthogonal to both $u$ and $v$) and $\theta$ is a scalar then

$$(1 + uv)\, e^{u\theta} = e^{v\theta}\, (1 + uv)$$

and

$$(1 - uv)\, e^{u\theta} = e^{-v\theta}\, (1 - uv)$$

*Multiplication Reordering Rule 4:* Given two unit vector quaternions $u$ and $v$ and scalars $\alpha$ and $\beta$

$$e^{u\alpha}\, e^{v\beta} = e^{w\beta}\, e^{u\alpha}$$

where

$$w = e^{u\alpha}\, v\, e^{-u\alpha}$$

$w$ is a precessed version of $v$, by angle $-2\alpha$ around axis $u$.

## 4 Quaternions representing light

As covered in Section 2, it is already known that a quaternion is a good way to represent an optical waveplate component. The electric field of a single beam of light, such as one transverse mode in an optical waveguide, can also be represented by a quaternion. Then the two systems come together and the problem of transmitting light through a series of waveplates can be written wholly in terms of quaternions. The usual way to express electric field is as a Jones vector $(E_x,E_y)^{\mathrm{T}}$ (the superscript T means vector transpose). $E_x$ and $E_y$ are complex numbers. To convert to a quaternion, the four quaternion elements can be associated with the Jones vectors corresponding to the left hand columns of (3). So Jones vector $(E_x,E_y)^{\mathrm{T}}$ becomes quaternion $q$, where

$$q = \mathrm{Re}(E_x) + \mathrm{Im}(E_x)\, i + \mathrm{Re}(E_y)\, j + \mathrm{Im}(E_y)\, k$$

If we consider $E_x$ and $E_y$ as complex numbers based on imaginary number $i$, so $i$ is like the existing imaginary number associated with a phase shift and $j$ and $k$ are new, then

$$q = E_x + E_y\, j \tag{8}$$

Table 2 gives examples of quaternions and the optical signals they represent.

**Table 2** Some optical signal cases and their quaternions

| quaternion | optical signal |
|---|---|
| 1 | horizontally polarized, zero phase |
| $i$ | horizontally polarized with a $\pi/2$ phase shift |
| $j$ | vertically polarized, zero phase |
| $k$ | vertically polarized with a $\pi/2$ phase shift |
| $1+j$ | linearly polarized at $\pi/4$, zero phase |
| $1+k$ | left circular polarized |
| $1-k$ | right circular polarized |

A quaternion is a natural way to represent a polarized optical signal. We are accustomed to using complex numbers to represent rotations. An attempt to express rotation of an electric field in physical space by multiplying by a complex number in *i* fails, because we already use imaginary number *i* to encode a phase shift of the electric field. What is needed is a different imaginary number, *j*, to express rotation in the transverse plane in physical space, and this is what the quaternion offers. As Hamilton discovered, a third imaginary number $k = ij$ is needed to make the algebra complete.

Since the transformation of an optical signal by a waveplate is computed via a Jones vector multiplied by a Jones matrix, it follows that the same transform may be represented wholly by quaternions. Optical field *q* is transformed to field *r* by waveplate *p*, where

$$r = qp$$

*p* is applied by right multiplication because, as stated in Section 1.2, adopting matrix basis set (3) leads to right multiplication for successive optical elements. An immediate advantage of using quaternions over Jones vectors/matrices is that the problem of computing waveplate *p* for a given *q* and *r* can be solved by quaternion division.

$$p = q\backslash r$$

It is not possible to get the answer directly from matrix calculus. The equivalent Jones expression would be

$$R = PQ$$

where *Q* and *R* are the input and output Jones vectors and *P* is the Jones matrix of the transforming optical device. There is no division process defined between two vectors to produce a matrix. That is because there are many matrices *P* that are solutions, which in turn is because there are many optical devices that transform from Jones vector *Q* to *R*. One of the matrix solutions *P* represents a waveplate (the desired answer), and the others are partial polarizers.

## 4.1 Phase shift & orthogonal polarization

The mathematical operation to apply a phase shift $\phi$ to the individual field components $E_x$ and $E_y$, which will again be considered as complex numbers in *i*, is

$$E_x \rightarrow E_x e^{i\phi} \qquad E_y \rightarrow E_y e^{i\phi}$$

Substituting into (8) gives the rule to apply the phase shift to quaternion $q$ representing the signal

$$q \to e^{i\phi} q$$

It is shown in Appendix B that $jq$ is orthogonal in polarization to $q$. Any phase shifted version of $jq$ shares that property, so the full set of states orthogonal to $q$ is $e^{i\phi}jq$, which includes the state $kq$.

The two operations in this section involve left multiplication of the quaternion $q$, to mathematically produce a phase shift or the orthogonal SOP. Transmission through a waveplate is represented by right multiplication, so it is not possible to apply a specified optical phase shift or to go to the orthogonal SOP by passing through a certain waveplate for all possible input states $q$. For a given $q$, however, there is a waveplate which produces the desired effect. For example, the set of waveplates $p$ to go to from $q$ to an orthogonal state satisfy

$$qp = e^{i\phi} jq$$

which has solution

$$p = q \backslash e^{i\phi} jq$$

## 4.2 Orthogonality conditions

There are several kinds of orthogonality that could apply between optical signals represented by quaternions $p$ and $q$. Quaternion-sense orthogonality has already been discussed in Section 3.1. The two signals could be orthogonal in state of polarization, or they could be orthogonal in phase, that is $\pi/2$ out of phase. The orthogonality status is encoded in the product $pq^{\dagger}$:

- if the two states are quaternion-sense orthogonal, $(pq^{\dagger})_0 = 0$
- if the two states have orthogonal SOPs, $(pq^{\dagger})_0 = (pq^{\dagger})_1 = 0$
- if the two states have the same SOP, $(pq^{\dagger})_2 = (pq^{\dagger})_3 = 0$
- if the two states have the same SOP and are orthogonal in phase, $(pq^{\dagger})_0 = (pq^{\dagger})_2 = (pq^{\dagger})_3 = 0$

## 4.3 Stokes vector & Poincaré sphere

The polarization state of an optical signal is indicated by a point on the Poincaré sphere. Linear states of polarization belong on the equator on the sphere, the poles identify the two circular states, and the rest of the sphere is for the more general elliptical states of polarization. The representation can be extended into a 3-dimensional vector, often known as the Stokes vector, where the vector's length is the magnitude squared of the signal's electric field. The Stokes vector's components in terms of electric field components $E_x$ and $E_y$ are [2]

$$\begin{aligned} S_1 &= |E_x|^2 - |E_y|^2 \\ S_2 &= 2\,\mathrm{Re}(E_x {E_y}^*) \\ S_3 &= 2\,\mathrm{Im}(E_x {E_y}^*) \end{aligned} \qquad (9)$$

The Stokes vector includes the SOP and magnitude but does not depend on the phase of the optical signal. Based on the quaternion algebra laid out already, the expression for $s$ below is reasonable as an attempt to convey the Stokes vector's information for a signal represented by quaternion $q$.

$$s = i\, q^{\dagger i}\, q \qquad (10)$$

We know the phase is adjusted by left multiplying $q$ (Section 4.1), so $q^{\dagger i}q$ has cancelled that phase via (6). $q^{\dagger i}q$ has zero $i$-component (Section 3.4), and multiplying by $i$ rearranges the coefficients to change it into a vector quaternion (zero scalar component).

Substituting (8) into (10), expanding, applying Multiplication Reordering Rule 2, and then comparing the result with (9) reveals that

$$s_1 = S_1 \qquad s_2 = S_3 \qquad s_3 = S_2$$

$s$ is indeed the same as the Stokes vector, aside from the rearrangement of coefficients 2 and 3. The quantity calculated in (10) will be known as the Stokes vector quaternion in this paper.

The definition of $s$ (10) offers an explanation why the angle of orientation and angle of ellipticity appear doubled on the Poincaré sphere compared to real space. Multiplying $q^{\dagger i}$ by $q$ cancels the phase, but squares the polarization part, and so it doubles the polarization-related exponents.

## 4.4 Polarization ellipse

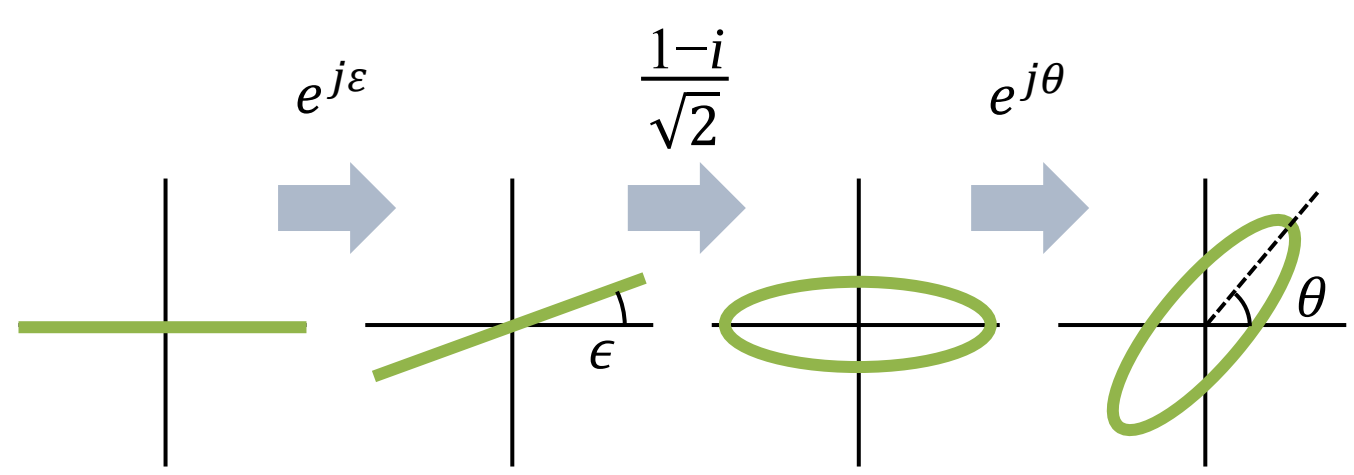

**Fig. 2.** Transformations to generate an elliptical SOP having angle of ellipticity $\varepsilon$ and orientation $\theta$. The axes in each of the four diagrams are the $x$ (horizontal) and $y$ (vertical) axes, and the green line is the electric field trajectory.

The polarization ellipse is a 2-D plot showing the trajectory mapped out by the electric field vector in the plane orthogonal to the direction of propagation. The ellipse is characterized by the angle of ellipticity $\varepsilon$ (the arctangent of the ratio of minor and major axes), which ranges from 0 (linear SOP) to $\pm\pi/4$ (circular), and by the angle of orientation $\theta$ of the major axis. In addition the signal is characterized by the optical phase $\phi$, and the field magnitude $R$.

Next, the quaternion algebra presented in Sections 1.2 and 3 will be used to calculate the quaternion $q$ associated with a signal having polarization ellipse parameters $\varepsilon$, $\theta$, $\phi$. The process shown in Fig. 2 is followed. The optical phase is only important as a relative quantity, and, for reasons which will become apparent, we start with a horizontally polarized signal, magnitude $R$, having phase $\phi+\pi/4$. The horizontal light is rotated by $\varepsilon$, and sent through a quarter waveplate so it becomes elliptically polarized with angle of ellipticity $\varepsilon$. Finally the elliptical state is rotated by physical angle $\theta$. The quaternion value of the optical signal after these steps is

$$q = R\, e^{i\left(\phi+\frac{\pi}{4}\right)}\, e^{j\varepsilon}\, \frac{1-i}{\sqrt{2}}\, e^{j\theta}$$

Applying Multiplication Reordering Rule 3 to the 3rd and 4th terms, and then cancelling the 2nd and 3rd terms which have parallel vector parts, gives

$$q = R\, e^{i\phi}\, e^{k\varepsilon}\, e^{j\theta} \tag{11}$$

The conclusion is that the quaternion representing an optical signal state is simply the product of the magnitude and the exponentials of the three polarization ellipse parameters.

In general a unit quaternion can be expressed as a product of exponentials of *i*, *j* and *k* in any order. Only if the exponentials are in the same order as (11), *i*-*k*-*j*, will the three exponential phases correspond to $\phi$, $\varepsilon$ and $\theta$.

As a further exercise in quaternion algebra, it will be shown how to calculate the polarization ellipse parameters starting from $q$. The ranges of the parameters are known to be

$$R \geq 0 \qquad -\pi < \phi \leq \pi \qquad -\frac{\pi}{2} < \theta \leq \frac{\pi}{2} \qquad -\frac{\pi}{4} \leq \varepsilon \leq \frac{\pi}{4}$$

First,

$$R = \sqrt{q^{\dagger} q}$$

Substituting (11) into (10),

$$s = iRe^{j\theta}e^{k\varepsilon}e^{-i\phi}e^{i\phi}e^{k\varepsilon}e^{j\theta}R = R^2 ie^{j\theta}e^{2k\varepsilon}e^{j\theta} \tag{12}$$

Taking the *j*-conjugate and applying Multiplication Reordering Rule 2,

$$s^{\dagger j} = R^2 e^{-j\theta}e^{2k\varepsilon}e^{-j\theta}i = R^2 ie^{j\theta}e^{-2k\varepsilon}e^{j\theta} \tag{13}$$

Subtracting (13) from (12) and applying Multiplication Reordering Rule 2 again

$$s - s^{\dagger j} = R^2 ie^{j\theta}2k \sin 2\varepsilon\, e^{j\theta} = R^2 ike^{-j\theta}e^{j\theta}2 \sin 2\varepsilon$$

Therefore

$$\sin 2\epsilon = -\frac{s_2}{R^2}$$

Next, to solve for $\theta$, add (12) and (13)

$$s + s^{\dagger j} = R^2 ie^{j\theta}2 \cos 2\varepsilon\, e^{j\theta} = R^2 ie^{j2\theta}2 \cos 2\varepsilon$$

Noting that $\cos 2\varepsilon > 0$ (except for the case of $\cos 2\varepsilon = 0$ which will be deferred),

$$\tan 2\theta = \frac{s_3}{s_1}$$

Given $\varepsilon$ and $\theta$ are known, $\phi$ is quickly obtained by substituting those values back into (11).

When $\cos 2\varepsilon = 0$, $\varepsilon = \pm\pi/4$, which corresponds to the two circular polarization states. Ellipse orientation $\theta$ is not defined for these states. In quaternion terms, we see that for right circular light, for example, the following equation is true for any value of scalar $\Delta$, by Multiplication Reordering Rule 3.

$$e^{i\phi}\ \frac{1-k}{\sqrt{2}}\ e^{j\theta} = e^{i(\phi+\Delta)}\ \frac{1-k}{\sqrt{2}}\ e^{j(\theta-\Delta)}$$

The quaternion theory is telling us, correctly, that for the circular SOPs there is ambiguity between $\phi$ and $\theta$.

## 5 Quaternion representation of a waveplate

The quaternion was shown in Section 2 to be a way to represent a waveplate component. The quaternion was a direct replacement for the Jones matrix describing a member of the closed set of waveplate transformations. In this section the quaternion will be derived for a specific waveplate.

Consider the waveplate of retardance $2\eta$, where unit quaternion $q$ represents the slow axis and $jq$ the fast axis. The component acts on the slow axis to increase its phase by $\eta$ and leave the SOP unchanged, so

$$qp = e^{i\eta}q$$

which has solution

$$p = q^{\dagger}e^{i\eta}q \tag{14}$$

We can see this is the correct expression for $p$. When the input is one of the quaternions lying on the slow axis $e^{i\phi}q$, the output is (making use of Multiplication Reordering Rule 1)

$$e^{i\phi}qp = e^{i\eta}e^{i\phi}q$$

For all $e^{i\phi}q$ the phase has been advanced by $\eta$. When the input is one of the states on the fast axis $e^{i\phi}jq$,

$$e^{i\phi}jqp = e^{-i\eta}e^{i\phi}jq$$

(using Multiplication Reordering Rules 1 & 2). The phase has been reduced by $\eta$. Next, applying Multiplication Reordering Rule 4 to (14)

$$p = e^{s\eta}q^{\dagger}q \tag{15}$$

where

$$s = q^{\dagger}iq \tag{16}$$

Simplifying (15) and substituting (5) into (16) gives

$$\begin{gathered} p = e^{s\eta} \\ s = i\,q^{\dagger i}\,q \end{gathered} \tag{17}$$

$s$ is recognizable as the Stokes vector quaternion of $q$, from equation (10). So a quaternion exponential represents a waveplate, where the unit vector part is the Stokes unit vector quaternion of the slow axis, and the angle is half the retardance.

The Stokes vector quaternion was derived in Section 4.3 as an equivalent to the familiar Stokes vector on the Poincaré sphere. In fact this quaternion appears centrally in the quaternion algebra of a waveplate, and, as will be seen in the next section, the algebra of a polarizer.

We can now recognize the vector $W$ of Section 2 as the logarithm of the quaternion associated with the waveplate. The precession around $W$ is explained in terms of the quaternion math of Section 3.3. The waveplate output quaternion $r$ for input $q$ is

$$r = q\,e^{s\eta}$$

so the Stokes vector quaternion of the waveplate output is

$$i\,r^{\dagger i}\,r = i\,(e^{s\eta})^{\dagger i}\,q^{\dagger i}\,q\,e^{s\eta}$$

Applying (5) gives

$$i\,r^{\dagger i}r = e^{-s\eta}\,i\,q^{\dagger i}\,q\,e^{s\eta}$$

According to Section 3.3 this is the equation for precession of the input Stokes vector by angle $2\eta$ around the fast/slow axis $s$, which is the behavior expected from Section 2. Equation (17) has connected the quaternion representation of the waveplate to its geometric representation in Stokes space.

## 6 Quaternion representation of a polarizer

As we have seen, the quaternion inherently represents an optical waveplate component, but a single quaternion cannot represent a polarizer. It is not possible to right multiply an optical signal by a certain quaternion and obtain a result which is the same as transmitting that optical signal through a polarizer. It is possible, however, to capture the effect of a polarizer by a more complicated algebraic expression.

Consider a partial polarizer (a polarization dependent loss element) for which unit quaternion $p$ lies on the high transmission SOP, and therefore $jp$ on the low transmission SOP. The field extinction of the partial polarizer is scalar $\mu$. The high transmission axis scales the electric field by 1, and the low transmission axis by $\mu$, where $0 \leq \mu \leq 1$. The output $r$ of the partial polarizer to input $q$ is

$$r = \tfrac{1}{2}\left((1+\mu)q - (1-\mu)q^{\dagger jk} is\right) \tag{18}$$

where

$$s = i\, p^{\dagger i}\, p$$

$s$ is the Stokes unit vector quaternion of the high transmission axis $p$. (18) can also be written, via (7), as

$$r = \frac{1}{2}\left((1+\mu)q - (1-\mu)iqs\right)$$

The expression (18) offers a clue how it acts as a polarizer. The $jk$-conjugate has two components inverted, so adding to the double conjugate eliminates two components and leaves behind the other two.

To verify that (18) is correct, set the input state $q$ to one of the high transmission SOP signals $e^{i\phi}p$.

$$r = \frac{1}{2}\left((1+\mu)e^{i\phi}p - (1-\mu)e^{i\phi}p^{\dagger jk} ii p^{\dagger i} p\right)$$

Recognizing $p^{\dagger jk}$ as the conjugate of $p^{\dagger i}$, so they cancel,

$$r = e^{i\phi}p = q$$

Then set $q$ to one of the quaternions representing the orthogonal state $e^{i\phi}jp$.

$$r = \frac{1}{2}\left((1+\mu)e^{i\phi}jp - (1-\mu)e^{i\phi}(-j)p^{\dagger jk} ii p^{\dagger i} p\right)$$

$$r = \mu e^{i\phi}jp = \mu q$$

Equation (18) does indeed produce the behavior of the specified partial polarizer.

Having added the polarizer component, we now know how to express the effect of all Jones matrices in terms of quaternions.

## 7 Endless phase shifter example

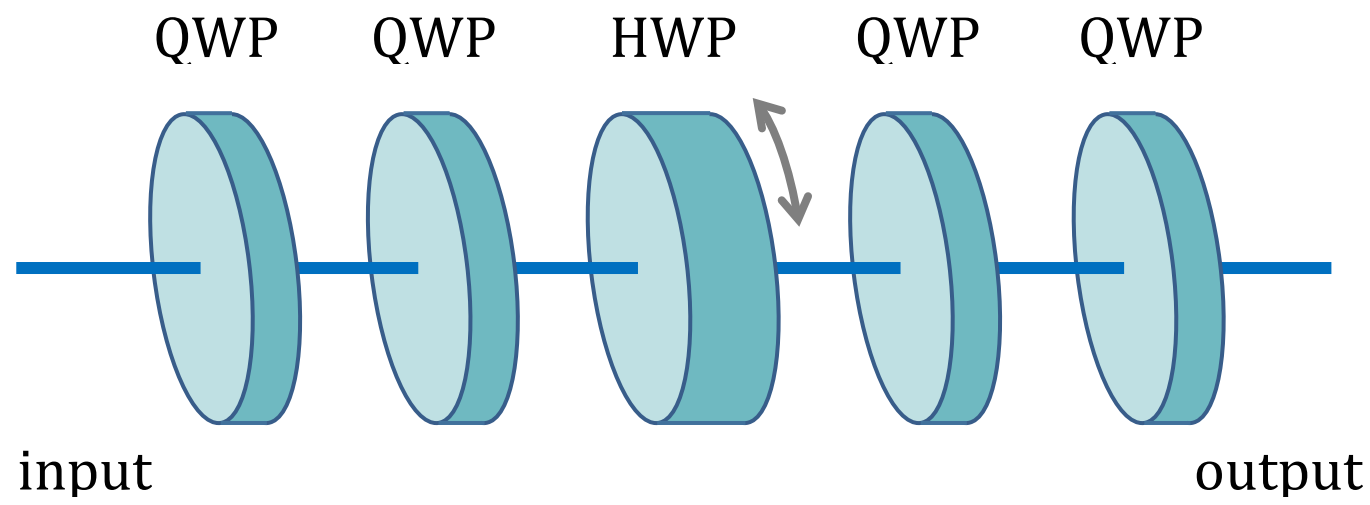

**Fig. 3.** The Evans phase shifter, comprising five waveplates. (QWP – quarter waveplate, HWP – half waveplate.)

The quaternion theory presented above will now be applied as an example to develop an application that has not been described before: an endless phase shifter comprising only three waveplates. A sequence of waveplates is often used to make a polarization controller. They usually fall into one of two configurations: a series of variable retardance waveplates that are at fixed orientations with respect to one another, or a series of fixed retardance waveplates that can be rotated. The latter option will be used here. It is already known that this kind of polarization controller can also produce an endless optical phase shift, via the configuration of Fig. 3, first described by Evans [19, 20]. The half waveplate at the center of the Evans phase shifter has a circular polarization state at its input, and a circular state (of the opposite sense) at its output. In general the rotation of a half wave plate produces a rotation of a signal's polarization ellipse. When it acts on a circular SOP it is equivalent to changing the phase of the optical signal, while leaving the output SOP unchanged. A rotation of the half waveplate by angle $\alpha$ produces a phase shift $2\alpha$. The two quarter waveplates to the left of the central half waveplate in Fig. 3 transform an arbitrary input SOP to a circular state, and the two quarter waveplates to the right transform the output circular state to an arbitrary output SOP. The Evans phase shifter therefore includes five waveplates.

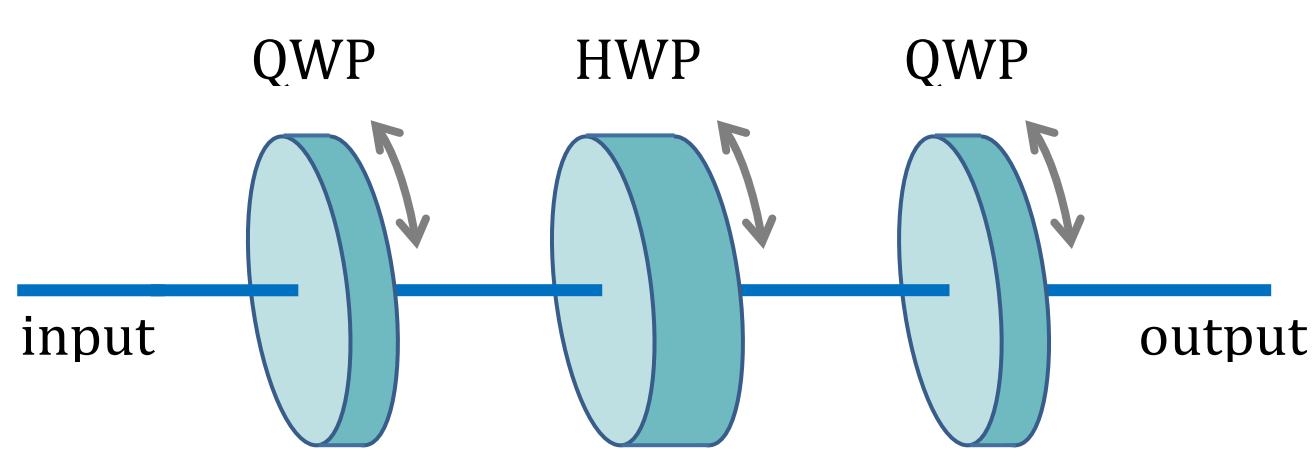

**Fig. 4.** Alternative phase shifter configuration using three waveplates. (QWP – quarter waveplate, HWP – half waveplate.)

The alternative configuration of Fig. 4 is capable of producing the same effect as the Evans phase shifter, transforming from a fixed arbitrary input to a fixed arbitrary output SOP and endlessly changing the signal's phase. It comprises three waveplates, having angles of orientation $\psi_a$, $\psi_b$, $\psi_c$, and this time all three rotate to produce the required phase shift. The waveplates are quarter-half-quarter waves. (In fact any order of these three waveplates will work, but only the central half waveplate configuration is studied here.)

In quaternion terms, transformation $p$ goes from input $q$ to output $r$ (both unit quaternions) and $r$ is phase shifted by $\phi$, so

$$p = q^{\dagger}\, e^{i\phi}\, r \tag{19}$$

which by Multiplication Reordering Rule 4 is the same as

$$p = e^{s\phi}\, q^{\dagger}\, r$$

where $s = iq^{\dagger i}q$ is the Stokes vector quaternion associated with $q$. Any value of unit quaternion $q^{\dagger}r$ and unit vector quaternion $s$ may be encountered.

The three waveplates of Fig. 4 have quaternion transformation

$$p = \tfrac{1}{2}\, e^{-j\psi_a}\,(1+i)\, e^{j\psi_a}\, e^{-j\psi_b}\, i\, e^{j\psi_b}\, e^{-j\psi_c}\,(1+i)\, e^{j\psi_c}$$

Expanding the (1+$i$) terms and applying Multiplication Reordering Rules 1 & 2

$$2p = ie^{j2\psi_b} - e^{j(2\psi_b-2\psi_a)} - e^{j(2\psi_c-2\psi_b)} - ie^{j(2\psi_a-2\psi_b+2\psi_c)}$$

The first and fourth terms on the RHS have $i$- and $k$-components only, and the second and third terms have only scalar and $j$-components. The quaternion equation separates into two simultaneous complex number equations in $j$.

$$2(p_0 + p_2 j) = -e^{j(2\psi_b-2\psi_a)} - e^{j(2\psi_c-2\psi_b)} \tag{20}$$

$$2(p_1 + p_3 j) = e^{j2\psi_b} - e^{j(2\psi_a-2\psi_b+2\psi_c)} \tag{21}$$

Factoring $e^{j(\psi_c-\psi_a)}$ from (20) and $e^{j(\psi_a+\psi_c)}$ from (21)

$$p_0 + p_2 j = -e^{j(\psi_c-\psi_a)}\,\cos(-\psi_a + 2\psi_b - \psi_c) \tag{22}$$

$$p_1 + p_3 j = j\, e^{j(\psi_a+\psi_c)}\,\sin(-\psi_a + 2\psi_b - \psi_c) \tag{23}$$

So

$$e^{j2\psi_a} = \pm\,\frac{j\,(p_1 + p_3 j)}{p_0 + p_2 j}\sqrt{\frac{{p_0}^2 + {p_2}^2}{{p_1}^2 + {p_3}^2}} \tag{24}$$

$$e^{j2\psi_c} = \pm\,\frac{j\,(p_0 + p_2 j)\,(p_1 + p_3 j)}{\sqrt{({p_0}^2 + {p_2}^2)\,({p_1}^2 + {p_3}^2)}} \tag{25}$$

There is a ± sign because the quadrant of $-\psi_a+2\psi_b-\psi_c$ has not yet been established. Substituting the square of (23) into (21) gives a quadratic equation in $e^{j2\psi_b}$, having solutions

$$e^{j2\psi_b} = (p_1 + p_3 j)\left(1 \pm j\sqrt{\frac{{p_0}^2 + {p_2}^2}{{p_1}^2 + {p_3}^2}}\right) \tag{26}$$

Choosing the consistent ± sign options from (24), (25) and (26), there are in general two waveplate angle solutions for quaternion transformation $p$

$$\psi_a = \frac{1}{2}\arg(p_1 + p_3 j) - \frac{1}{2}\arg(p_0 + p_2 j) + \frac{\pi}{4}$$

$$\psi_b = \frac{1}{2}\arg(p_1 + p_3 j) - \frac{1}{2}\arctan\sqrt{\frac{{p_0}^2 + {p_2}^2}{{p_1}^2 + {p_3}^2}}$$
$$\psi_c = \frac{1}{2}\arg(p_1 + p_3 j) + \frac{1}{2}\arg(p_0 + p_2 j) + \frac{\pi}{4}$$

(27)

and

$$\psi_a = \frac{1}{2}\arg(p_1 + p_3 j) - \frac{1}{2}\arg(p_0 + p_2 j) - \frac{\pi}{4}$$
$$\psi_b = \frac{1}{2}\arg(p_1 + p_3 j) + \frac{1}{2}\arctan\sqrt{\frac{{p_0}^2 + {p_2}^2}{{p_1}^2 + {p_3}^2}}$$
$$\psi_c = \frac{1}{2}\arg(p_1 + p_3 j) + \frac{1}{2}\arg(p_0 + p_2 j) - \frac{\pi}{4}$$

(28)

where, for equations (27) and (28) and all equations in this section that compute a waveplate angle of rotation $\psi$, the summation is evaluated modulo $\pi$ so that $-\frac{\pi}{2} < \psi \leq \frac{\pi}{2}$.

The derivation above fails when $p_0 = p_2 = 0$ or $p_1 = p_3 = 0$. It can be seen from (22) and (23) that in these cases a constraint on either $\psi_a - \psi_c$ or $\psi_a + \psi_c$ is removed, and $\psi_a$ and $\psi_c$ can take on a continuum of values. Solutions are obtained by following a similar procedure to the previous paragraph. When $p_0 = p_2 = 0$

$$\psi_a = \frac{1}{2}\arg(p_1 + p_3 j) + \frac{\pi}{4} + \alpha$$
$$\psi_b = \frac{1}{2}\arg(p_1 + p_3 j)$$
$$\psi_c = \frac{1}{2}\arg(p_1 + p_3 j) + \frac{\pi}{4} - \alpha$$

(29)

where $\alpha$ can take on any scalar value. When $p_1 = p_3 = 0$

$$\psi_b = \text{any value}$$
$$\psi_a = \psi_b - \frac{1}{2}\arg(p_0 + p_2 j) + \frac{\pi}{2}$$
$$\psi_c = \psi_b + \frac{1}{2}\arg(p_0 + p_2 j) + \frac{\pi}{2}$$

(30)

Given the optical phase, ellipse angle of orientation and angle of ellipticity are $\phi_{\text{in}}$, $\theta_{\text{in}}$, $\varepsilon_{\text{in}}$ for the input to the waveplate system and $\phi_{\text{out}}$, $\theta_{\text{out}}$, $\varepsilon_{\text{out}}$ for its output, the condition $p_0 = p_2 = 0$ corresponds to

$$\varepsilon_{out} = -\varepsilon_{in} \quad \text{and} \quad \phi_{out} = \phi_{in} \pm \frac{\pi}{2}$$

The condition $p_1 = p_3 = 0$ happens when

$$\varepsilon_{out} = \varepsilon_{in} \quad \text{and} \quad \phi_{out} = \phi_{in} + \frac{\pi}{2} \pm \frac{\pi}{2}$$

These states of the waveplate system are singular states. The Jacobian matrix from waveplate angle space $\psi_a$, $\psi_b$, $\psi_c$ to quaternion transformation space $p_1$, $p_2$, $p_3$, $p_4$ has one of its singular values equal to zero at these states.

The waveplate orientations to produce a phase shift $\phi$ can be calculated from either (27) or (28), where $p$ comes from (19). If the exact singular condition is encountered then there is a range of solutions (29) or (30). Thus, with only three waveplates there is a solution for all possible input signal states $q$, output states $r$, and phase shift $\phi$. As an example, Fig. 5 shows the waveplate angles computed from (27) to produce a phase ramp of $2\pi$ to a signal going from $q = -\frac{8}{9} + \frac{2}{9}i + \frac{1}{3}j + \frac{2}{9}k$ to $r = \frac{2}{7} - \frac{3}{7}i - \frac{6}{7}k$. Fig. 6 shows the effect of the waveplate system, expressed as the signal parameters of the output. The state of polarization is constant and the phase experiences a linear ramp of $2\pi$, which confirms that the equations developed in this section are correct.

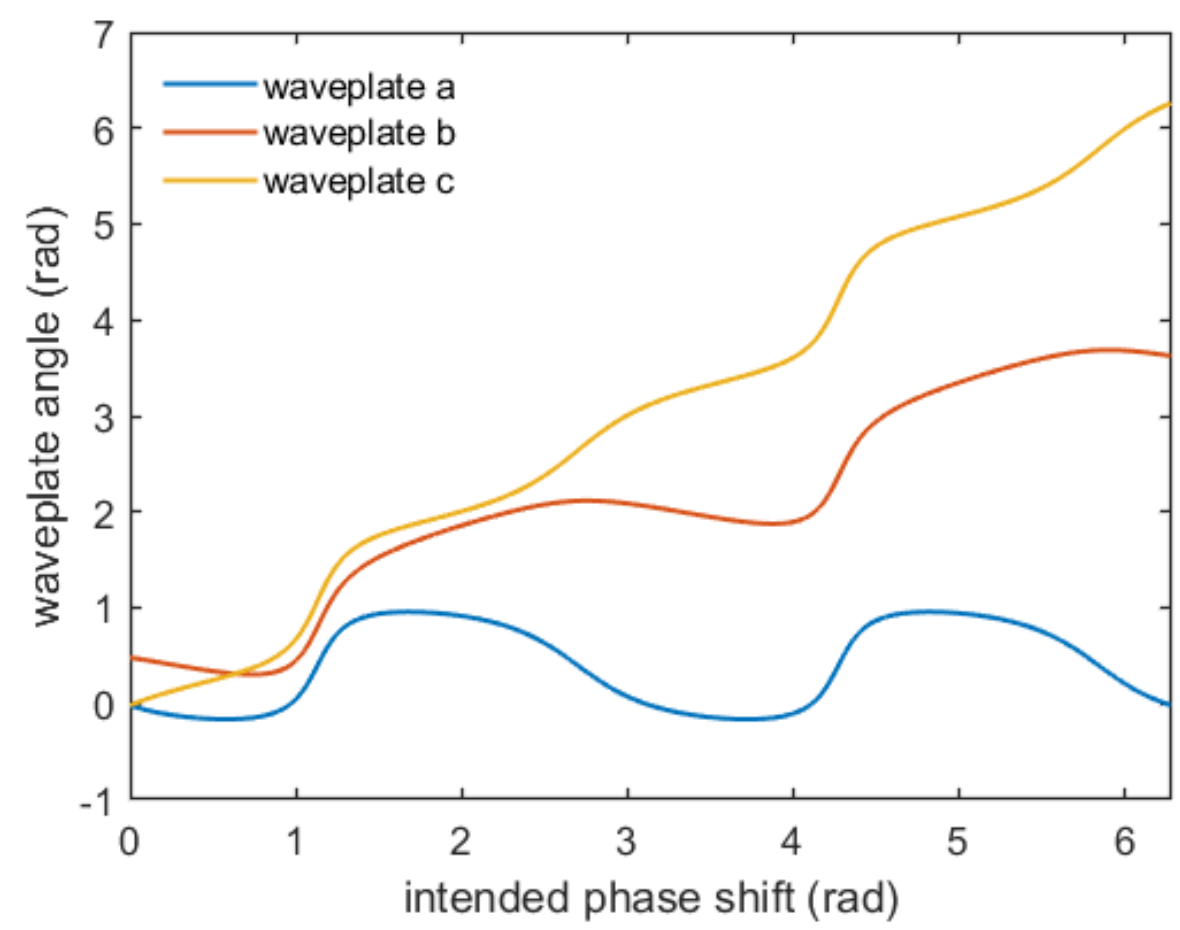

**Fig. 5.** Computed waveplate orientations to obtain $2\pi$ phase ramp for typical input & output optical states.

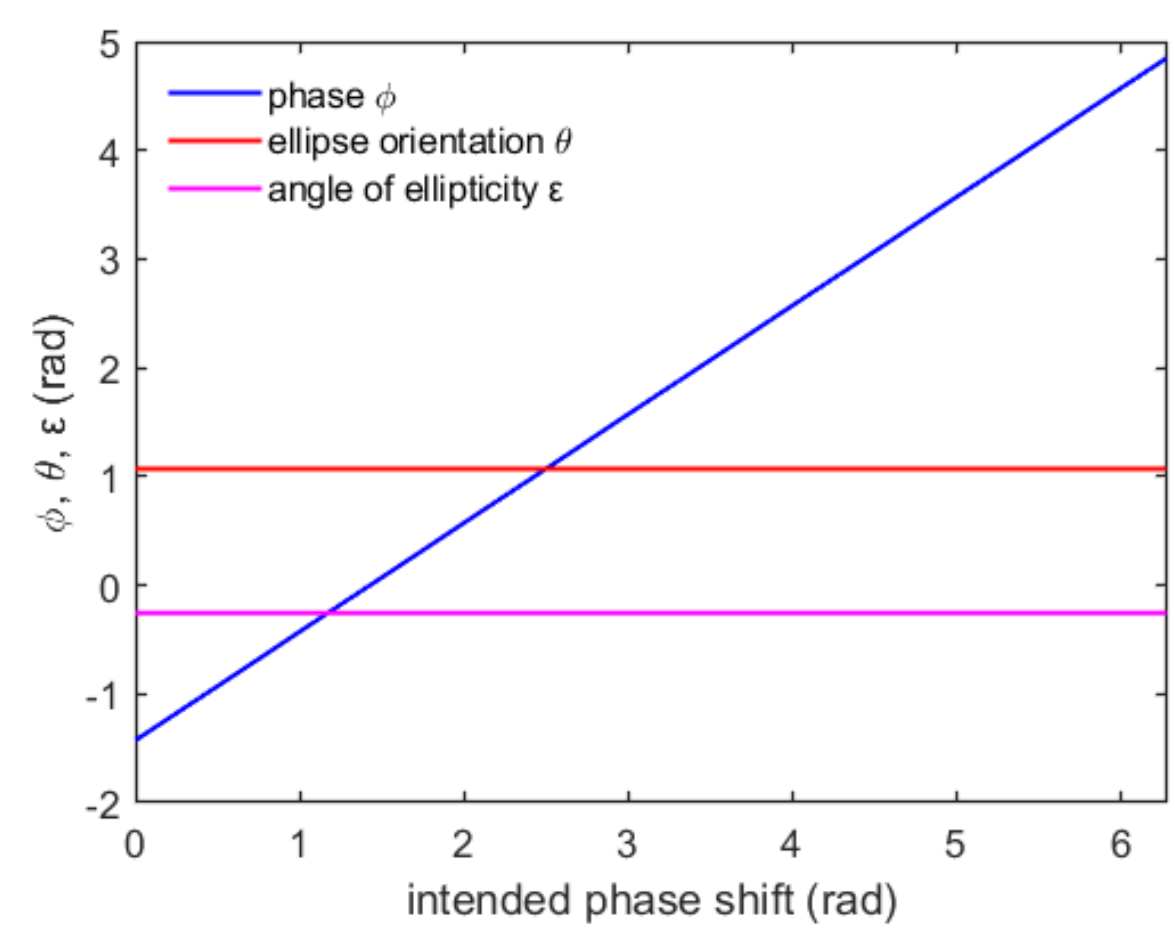

**Fig. 6.** Phase & polarization ellipse parameters of output optical signal for waveplate orientations of Fig. 5.

Also considered is the scenario where $q = -\frac{5}{6} + \frac{1}{6}i + \frac{1}{2}j + \frac{1}{6}k$ and $r = \frac{1}{3} - \frac{2}{3}i - \frac{2}{3}k$. These two states have the same angle of ellipticity (-0.23 rad), which means the waveplate system passes through two singular states during the $2\pi$ phase ramp. The waveplate rotation angles are shown in Fig. 7 and the resulting output state parameters in Fig. 8. The singularities are seen as large steps of $\pi/2$ in waveplate orientation, while the phase of the output signal experiences only a small increment. This large step in angle is similar to the unwinding event that is a feature of some polarization controller designs [21]. The waveplate system does produce the desired output when at or close to a singularity, but it clearly requires the bandwidth of the waveplate rotation subsystem to be much higher than the rate of phase change. In this regard the three waveplate solution is inferior to the standard Evans phase shifter of Fig. 3, which always requires 1 mrad of waveplate rotation for 2 mrad of phase shift.

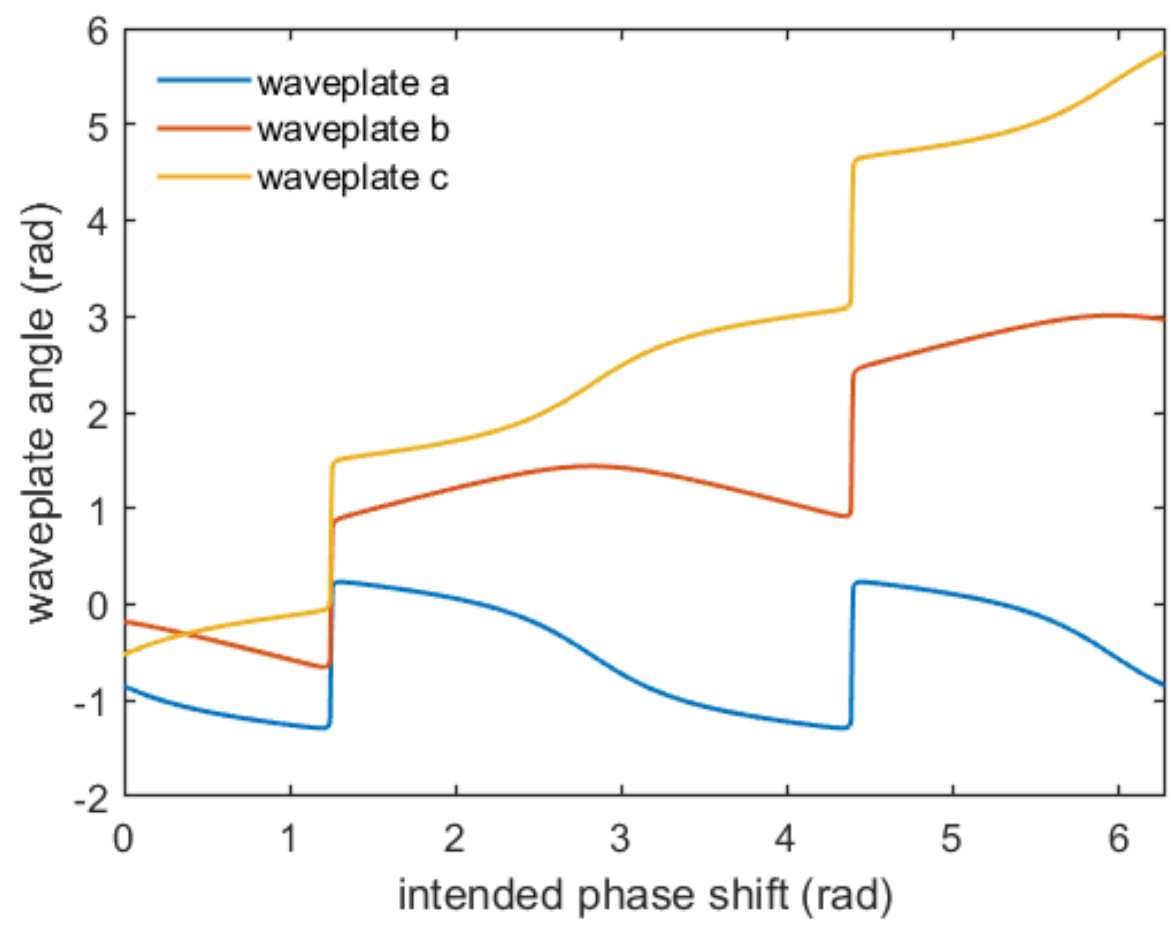

**Fig. 7.** Computed waveplate orientations to obtain $2\pi$ phase ramp for case where $\varepsilon_{out} = \varepsilon_{in}$.

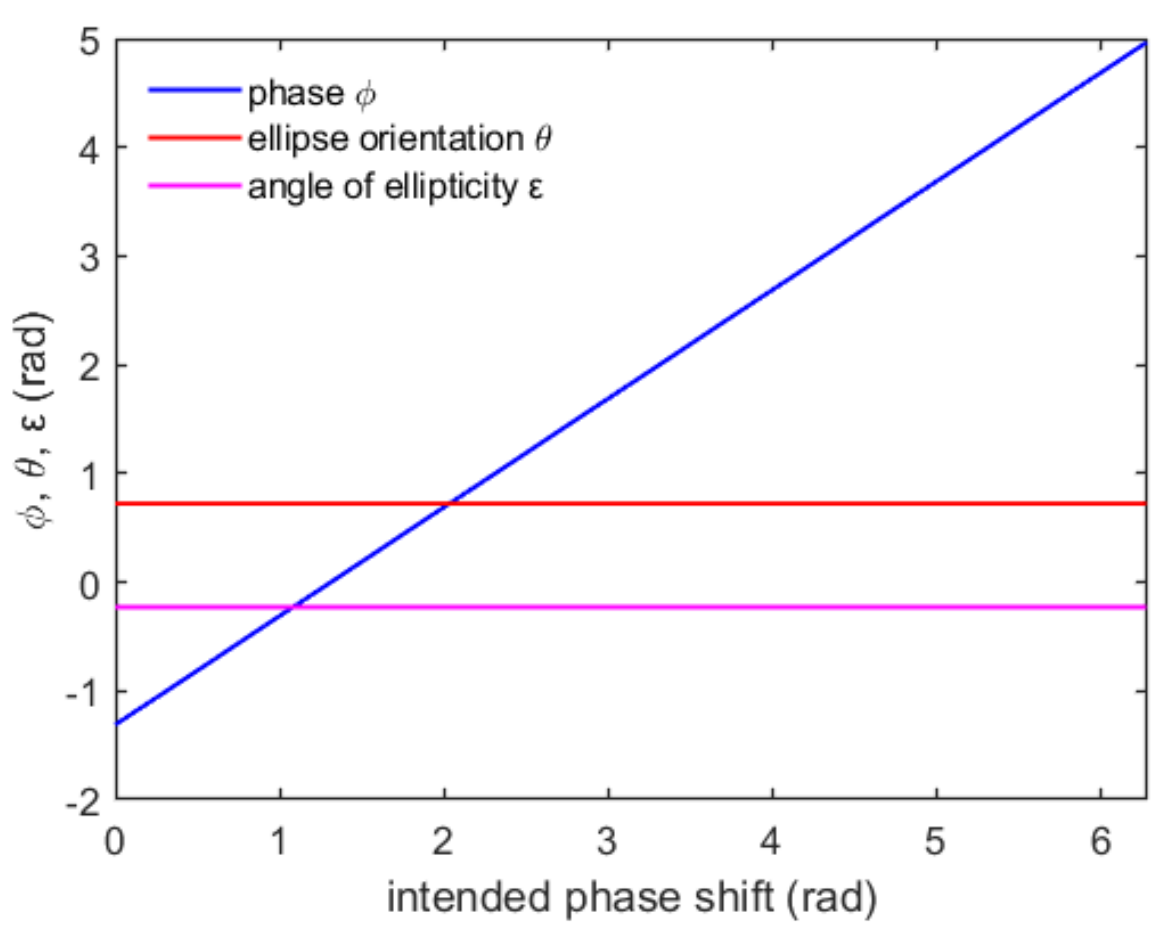

**Fig. 8.** Phase & polarization ellipse parameters of output optical signal for waveplate orientations of Fig. 7, the case where $\varepsilon_{out} = \varepsilon_{in}$.

The phase shifter was presented as an example of how the quaternion theory can be applied to capture the behavior of multiple waveplate devices and solve in the reverse direction, finding the device conditions to give the required output. In the past geometrical arguments have been used to solve such a system of waveplates (a polarization controller), based around the precession behavior seen on the Poincaré sphere [21]. If algebra is used instead, for example [22], it requires much effort as many variables are needed to define the waveplates and states of polarization. The analytical expressions to obtain the desired phase shift in this paper's solution have a compact form. Equations (27), (28), (29) and (30) are functions of quaternion $p$, which comes from the quaternion division of two polarization states (19). Stating these relations without using quaternions would take up much more space, exposing the many parameters of the waveplates and input and output SOPs. The quaternion approach has allowed us to solve using algebra a waveplate system that cannot be reached by geometrical arguments, but without the complexity unavoidable by starting from Jones vectors and Jones matrices.

## 8 Singularities in polarization controllers

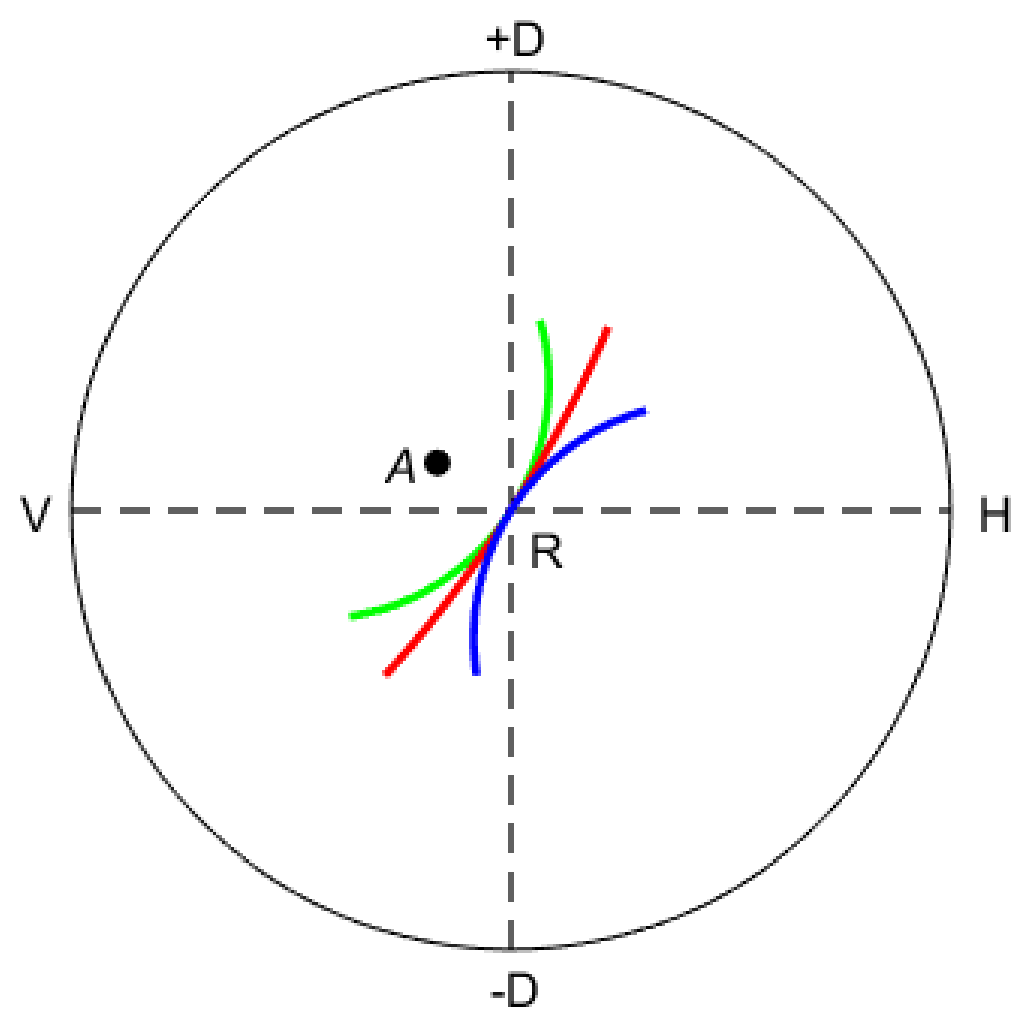

**Fig. 9.** Poincaré sphere showing effect on output SOP of dithering orientation of three waveplates in turn by ±0.1 rad; green – $\theta_a$ dithered, red – $\theta_b$, blue – $\theta_c$. Waveplates have arbitrarily chosen retardances (2.3, 2, 1.4 rad), and orientations are 0.5, 2.07, 2.07 rad, i.e. their fast/slow axes are aligned.

The second example of quaternions being applied is to a system of waveplates acting as a polarization controller rather than a phase adjuster. A sequence of waveplates is considered having fixed retardances that can be rotated to any angle of orientation. Such a polarization controller can fail in some circumstances when asked to continuously track a changing target. Fig. 9 shows the output of a three-waveplate polarization controller on the Poincaré sphere, where the current output state is right circular, at the centre of the diagram. Suppose we want the polarization state to move to nearby point $A$. The three curves show the effect on SOP of moving the three waveplates in turn by ±0.1 rad in orientation. The three curves are orthogonal to the direction from $R$ to $A$. No adjustment takes the output polarization closer to the target. In colloquial terms, the polarization controller is stuck.

What has happened is that the system of waveplates is in a singular state. In fact there is a set of adjustments that take the output SOP from $R$ to $A$, but the transition involves simultaneous large changes in waveplate orientations, similar to the situation with the singular state in Section 7. The standard algorithm to control the waveplate orientations is to dither them in turn while looking at the distance to the target, and then adjust the waveplate angles in the direction of most improvement [23]. That algorithm is frozen by the singularity, and when close to the singularity it becomes slow. There is some mention of singular states in polarization controllers in research papers (though it is not given that name). Walker & Walker [21] state that a system of fixed retardance waveplates does not provide endless polarization control even though the rotation of the waveplate elements is endless. Chen & Heismann [24] provide a more complete description of the phenomenon.

This section uses quaternions to present a mathematical explanation of the singular states, and identifies the conditions causing singular states and how often they occur. The quaternion is a good tool to use because the exponentials representing waveplates can be differentiated to provide the direction of change in output, and the quaternion rules described earlier allow the mathematical expressions to be manipulated easily. Let the input and output states be unit quaternions $q$ and $r$ and the polarization controller be represented by unit quaternion $p$, so once again

$$r = qp$$

The orientations of the waveplates are given by variable $\psi$ with a subscript, and the expression $\frac{\partial r}{\partial \psi}$ (where $\psi$ has no subscript) means the derivative w.r.t. any of the waveplate orientations $\psi$. We know from Section 4.1 that if

$$\frac{\partial r}{\partial \psi} = \alpha i r$$

where $\alpha$ is a scalar, then rotating that waveplate changes the phase of the optical signal. Similarly, if $\frac{\partial r}{\partial \psi} r^{\dagger}$ lies in the $jk$-plane, that is if

$$\frac{\partial r}{\partial \psi} r^{\dagger} = r\, p^{\dagger} \frac{\partial r}{\partial \psi} r^{\dagger} = \alpha e^{i\beta} j \tag{31}$$

($\beta$ is a scalar) then rotating that waveplate changes the SOP and does not change the phase. Note that since $p$ is a unit vector, $\frac{\partial p}{\partial \psi}$ is orthogonal to $p$, and $p^{\dagger} \frac{\partial p}{\partial \psi}$ is a vector quaternion. When the singularity occurs there is a certain direction on the $jk$-plane (a certain value of $\beta$) which is orthogonal to all the $\frac{\partial r}{\partial \psi} r^{\dagger}$. Recognizing that (31) is an example of quaternion precession (Section 3.3), it is clear that if all the the $p^{\dagger} \frac{\partial p}{\partial \psi}$ are orthogonal to a certain vector quaternion $w$, then a value of $r$ can be found that makes $rwr^{\dagger}$ lie in the $jk$-plane. Therefore the condition for a singularity is that there is a vector direction $w$ such that all the $p^{\dagger} \frac{\partial p}{\partial \psi}$ are orthogonal to $w$.

A polarization controller comprising three waveplates of half-retardance $\eta_a$, $\eta_b$, $\eta_c$ and angles of orientation $\psi_a$, $\psi_b$, $\psi_c$, has transformation

$$p = e^{-j\psi_a}\, e^{i\eta_a}\, e^{j\psi_a}\, e^{-j\psi_b}\, e^{i\eta_b}\, e^{j\psi_b}\, e^{-j\psi_c}\, e^{i\eta_c}\, e^{j\psi_c}$$

Differentiating a quaternion exponential is like differentiating a conventional exponential, but the product term must stay in place. Applying Multiplication Reordering Rules 1 and 2,

$$\frac{\partial p}{\partial \psi_a} = \left(-je^{-j\psi_a}e^{i\eta_a}e^{j\psi_a} + e^{-j\psi_a}e^{i\eta_a}je^{j\psi_a}\right)e^{-j\psi_b}\, e^{i\eta_b}\, e^{j\psi_b}\, e^{-j\psi_c}\, e^{i\eta_c}\, e^{j\psi_c}$$

$$\frac{\partial p}{\partial \psi_a} = 2\sin\eta_a\, e^{-j2\psi_a}\, e^{j\psi_b}\, e^{-i\eta_b}\, e^{-j\psi_b}\, e^{j\psi_c}\, e^{-i\eta_c}\, e^{-j\psi_c} k$$

and

$$p^\dagger \frac{\partial p}{\partial \psi_a} = 2\sin\eta_a\, e^{-j\psi_c}e^{-i\eta_c}e^{j\psi_c}e^{-j\psi_b}e^{-i\eta_b}e^{j\psi_b}e^{-j\psi_a}e^{-i\eta_a}e^{-j\psi_a}e^{j\psi_b}e^{-i\eta_b}e^{-j\psi_b}e^{j\psi_c}e^{-i\eta_c}e^{-j\psi_c}k$$

Similarly

$$p^\dagger \frac{\partial p}{\partial \psi_b} = 2\sin\eta_b\;\, e^{-j\psi_c}\, e^{-i\eta_c}\, e^{j\psi_c}\, e^{-j\psi_b}\, e^{-i\eta_b}\, e^{-j\psi_b}\, e^{j\psi_c}\, e^{-i\eta_c}\, e^{-j\psi_c}\, k$$

$$p^\dagger \frac{\partial p}{\partial \psi_c} = 2\sin\eta_c\;\, e^{-j\psi_c}\, e^{-i\eta_c}\, e^{-j\psi_c}\, k$$

We will study the condition where the waveplates are aligned with one another, fast axis to either fast or slow axis. That is

$$\psi_b = \psi_a + \frac{\pi}{4} \pm \frac{\pi}{4} \qquad\qquad \psi_c = \psi_a + \frac{\pi}{4} \pm \frac{\pi}{4}$$

First look at $\psi_a = \psi_b = \psi_c$. The three derivative equations become (via Multiplication Reordering Rules 1 & 2)

$$p^\dagger \frac{\partial p}{\partial \psi_a} = 2\sin\eta_a\;\, e^{-j\psi_a}\, e^{-i(\eta_a+2\eta_b+2\eta_c)}\, e^{-j\psi_a}\, k$$

$$p^\dagger \frac{\partial p}{\partial \psi_b} = 2\sin\eta_b\;\, e^{-j\psi_a}\, e^{-i(\eta_b+2\eta_c)}\, e^{-j\psi_a}\, k$$

$$p^\dagger \frac{\partial p}{\partial \psi_c} = 2\sin\eta_c\;\, e^{-j\psi_a}\, e^{-i\eta_c}\, e^{-j\psi_a}\, k$$

For the polarization controller states with $\frac{\pi}{2}$ rotations the three $p^\dagger \frac{\partial p}{\partial \psi}$ have the same form

$$p^\dagger \frac{\partial p}{\partial \psi} = 2\sin\eta_c\;\, e^{-j\psi_a}\, e^{i(\dots)}\, e^{-j\psi_a}\, k \tag{32}$$

but the exponent ( … ) is a different combination of $\eta_a$, $\eta_b$, $\eta_c$. Then if we consider systems with more than three waveplates, but keeping the fast/slow axes aligned with one another, it can be seen that all the $p^\dagger \frac{\partial p}{\partial \psi}$ still have the form of (32). All these $p^\dagger \frac{\partial p}{\partial \psi}$ are orthogonal to a forbidden direction $w = i\, e^{j2\theta_a}$, since

$$\mathrm{Sc}\left(e^{-j\psi_a}\, e^{i(\dots)}\, e^{-j\psi_a}\, \left(i\, e^{j2\psi_a}\right)^\dagger\right) = 0$$

Therefore all the polarization controller states with fast/slow axes aligned are singular states. We can solve for the output signal *r* using (31). One solution is

$$r = \frac{1}{2}\sqrt{2}\, e^{i\gamma}\, (1 + k)$$

where $\gamma$ can take on any scalar value, which is a solution of (31) given $\alpha = 1, \beta = 2\psi_a + 2\gamma$. Another solution is

$$r = \frac{1}{2}\sqrt{2}\, e^{i\gamma}\, (1 - k)$$

These $r$ are the left and right circular states, so the singularity is seen when the output SOP is circular.

The conclusion of this section is that all polarization controller designs comprising fixed retardance rotatable waveplates exhibit several singular states, no matter how many waveplate elements and what the retardances of the waveplates are. The singular states cause the polarization controller to stop working. The singularities happen when the fast and slow axes of the waveplates are aligned with one another, and when the output SOP is circular. Increasing the number of waveplates makes the singularity event less likely to be encountered.

## Appendix

### A. Other basis sets

The four matrices (3) are not the only basis set that can produce the quaternion multiplication rules (1) for a given multiplication direction (right-to-left or left-to-right). Hidayat [9] employed the following basis set.

$$1 \to \begin{pmatrix} 1 & 0 \\ 0 & 1 \end{pmatrix} \qquad i \to \begin{pmatrix} -h & 0 \\ 0 & h \end{pmatrix}$$
$$j \to \begin{pmatrix} 0 & -h \\ -h & 0 \end{pmatrix} \qquad k \to \begin{pmatrix} 0 & -1 \\ 1 & 0 \end{pmatrix}$$

This set does not have $i$ and $j$ corresponding to the intuitive quarter turn in phase and in physical space respectively.

Richartz & Hsu [3] used a basis set closer to this paper.

$$1 \to \begin{pmatrix} 1 & 0 \\ 0 & 1 \end{pmatrix} \qquad i \to \begin{pmatrix} h & 0 \\ 0 & -h \end{pmatrix}$$
$$j \to \begin{pmatrix} 0 & 1 \\ -1 & 0 \end{pmatrix} \qquad k \to \begin{pmatrix} 0 & h \\ h & 0 \end{pmatrix}$$

In fact these matrices are the transpose of the matrices in (3). This basis set does not lead to the mapping from electric field to quaternion in the very simple form of (8), however.

### B. Proofs of theorems

Most of the quaternion theorems presented in Section 1.2 and Section 3 are readily proved by substituting into the definition of a quaternion $q$ or the expression for product $pq$ in Section 1.2. The proof of Section 3.3 is given in chapter 5 of [5].

To derive (5) in Section 3.4, note that $\mathrm{Sc}(q^\dagger v)$ is the scalar product of quaternion $q$ with unit vector quaternion $v$. The component of $q$ in the direction of $v$ is $\mathrm{Sc}(q^\dagger v)v$, so

$$q^{\dagger v} = q - 2\,\mathrm{Sc}\big(q^\dagger v\big)\, v$$
$$q^{\dagger v} = q - v\big(q^\dagger v - vq\big)$$

$$q^{\dagger v} = q - vq^{\dagger}v + v^2 q$$
$$q^{\dagger v} = -vq^{\dagger}v$$

The other results of Section 3.4 are quickly derived from (5).

The first three Multiplication Reordering Rules of Section 3.5 are proved by substituting into the expression for a quaternion product in Section 1.2. In the case of Multiplication Reordering Rule 3, first break the exponential into its cosine and sine terms according to Euler’s theorem. For Multiplication Reordering Rule 4 the $e^{v\beta}$ and $e^{w\beta}$ exponentials are separated into sine and cosine terms

$$e^{u\alpha}\,(\cos\beta + v\sin\beta) = (\cos\beta + w\sin\beta)\,e^{u\alpha}$$
$$\sin\beta\,e^{u\alpha}\,v = \sin\beta\,w\,e^{u\alpha}$$

which is indeed true if $w = e^{u\alpha}\,v\,e^{-u\alpha}$.

To prove that $e^{i\phi}jq$ is orthogonal in SOP to $q$ (Section 4.1), calculate the Stokes vector quaternion of $e^{i\phi}jq$.

$$i\left(e^{i\phi}\,j\,q\right)^{\dagger i} e^{i\phi}\,j\,q$$
$$i\,q^{\dagger i}\,j\,e^{-i\phi}\,e^{i\phi}\,j\,q$$
$$-i\,q^{\dagger i}\,q$$

The Stokes vector quaternion of $e^{i\phi}jq$ is the negative that of $q$.

The orthogonality conditions of Section 4.2 are readily proved by substituting in $p = Re^{i\phi}jq$ for orthogonal SOPs, $p = Re^{i\phi}q$ for the same SOP, and $p = \pm Riq$ for the same SOP with orthogonal phase, where $R$ is a scalar.

## Acknowledgment

The author thanks J. E. Taylor for introducing him to the topic of quaternions.